# Monitoring the Solar Radius from the Royal Observatory of the Spanish Navy during the Last Quarter–Millennium

J.M. Vaquero[1,2] • M.C. Gallego[2,3] • J.J. Ruiz-Lorenzo[2,3] • T. López-Moratalla[4] • V.M.S. Carrasco[2,3] • A.J.P. Aparicio[2,3] • F.J. González-González[4] • E. Hernández-García[5]

[1] Departamento de Física, Universidad de Extremadura, Mérida (Badajoz), Spain, jvaquero@unex.es

[2] Instituto Universitario de Investigación del Agua, Cambio Climático y Sostenibilidad (IACYS), Universidad de Extremadura, Badajoz, Spain

[3] Departamento de Física, Universidad de Extremadura, Badajoz, Spain

[4] Real Instituto y Observatorio de la Armada, San Fernando (Cádiz), Spain

[5] Departamento de Matemáticas, Universidad de Extremadura, Mérida (Badajoz), Spain

**Abstract.** The solar diameter has been monitored at the Royal Observatory of the Spanish Navy (today the *Real Instituto y Observatorio de la Armada*: ROA) almost continuously since its creation in 1753 (*i.e.* during the last quarter of a millennium). After a painstaking effort to collect data in the historical archive of this institution, we present here the data of the solar semidiameter from 1773 to 2006, making up an extensive new database for solar-radius measurements can be considered. We have calculated the solar semidiameter from the transit times registered by the observers (except values of the solar radius from the modern Danjon astrolabe, which were published by ROA). These data were analysed to reveal any significant long-term trends, but no such trends were found. Therefore, the data sample confirms the constancy of the solar diameter during the last quarter of a millennium (approximately) within instrumental and methodological limits. Moreover, no relationship between solar radius and the new sunspot-number index has been found from measurements of the ROA. Finally, the mean value for solar semidiameter (with one standard deviation) calculated from the observations made in the ROA (1773–2006), after applying corrections by refraction and diffraction, is equal to 958.87"±1.77".

# 1. Introduction

The Sun is the driver of numerous processes on Earth and our near-space environment. In addition, it is the nearest and best observed star. Therefore, the long-term evolution of the Sun is of great interest to astrophysicists and geophysicists. Furthermore, variations in the solar diameter would have effects on the climate of our planet. For instance, during a solar cycle, the solar-radius variations are very small (Bush, Emilio, and Kuhn, 2010; Meftah *et al.*, 2015). Despite this fact, small changes in the solar luminosity due to a possible secular decrease or increase of the solar radius could affect the Earth's climate (Sofia *et al.*, 1979; Haigh, 1996, Foukal *et al.*, 2006). Long time series that provide information about solar activity come from, among others, isotopic production due to cosmic rays (Usoskin, 2013), covering the last few millennia. However, time series related to direct observations of the Sun are shorter. The naked-eye sunspot observations cover the last two millennia (Vaquero, Gallego, and García, 2002; Vaquero and Vázquez, 2009), and the telescopic observation of sunspots covers the last four centuries (Vaquero, 2007; Clette *et al.*, 2014).

Another important set of direct solar observations is related to measurements of the solar diameter. The first telescopic devices to measure the solar diameter were conceived in the 17th century. Thus, we have today a long series of measurements of the solar diameter during the last four centuries (Rozelot and Damiani, 2012). However, the apparent solar-radius variations differ according to the instruments and methods used, even for the most recent determinations. The origin of these differences might be the Sun, the Earth's atmosphere (in the case of ground-based observations), the instrument, or even the observer (especially in historical observations). Therefore, the recovery of historical solar-radius measurements could shed more light on and suggest new approaches to this old problem (Sigismondi and Fraschetti, 2001; Meftah *et al*., 2014).

In addition, the long-term trend of the solar diameter is still an open and controversial issue. We can cite that Eddy and Boornazian (1979) showed from measurements of the solar diameter made at Greenwich during the period 1836–1953 a decrease of 2.25" per century. Furthermore, Gilliland (1981), after analyzing five different data sets, concluded that a secular decrease of ≈0.1" per century over the last 265 years is likely, and Ribes, Ribes, and Barthalot (1987) suggest from observations recorded by French astronomers Picard and La Hire that solar diameter was about 4" greater around 1700 than now. In contrast, other investigations conclude that no long-term trend in the solar diameter is detected. For example, Parkinson, Morrison, and Stephenson (1980) find that previous reports of contraction of the solar diameter can be attributed to misinterpretations of the meridian-circle observations. Other authors such as Wittman (1980), Morrison, Stephenson, and Parkinson (1988), and Toulmonde (1997) do not support the assumption of a secular decrease. Furthermore, some works examine the relationships between solar activity and variations in the solar diameter. For example, Gilliland (1981) and Laclare *et al*. (1996) concluded that variations in the values of the solar radius are negatively correlated with sunspot-number index while Noël (2004) shows positive correlations between both. Meanwhile, Gough (1988) analysed the connection between solar diameter, sunspot number, and solar neutrinos.

The aim of this work is to recover the measurements of the solar radius made at the Royal Observatory of the Spanish Navy (hereafter ROA) since its foundation in 1753, to provide a machine-readable version for the solar-astrometry community, and to offer a first analysis of this set of observations. Thereby, a new set of solar radius observations over the last quarter of a millennium is available to evaluate the variations of solar diameter.

The ROA, located originally in the city of Cádiz, started its astronomical observations in 1753 (Lafuente and Sellés, 1988). The Observatory was transferred in 1798 to the nearby town of La Isla de León, which is today called San Fernando (González-González, 1992, 2004). The first (last) preserved measurements of the solar radius are dated in 1773 (2006). Today, the library and the archive of this institution preserve much of the heritage accumulated over the centuries, including archival material (González-González, 2010), a remarkable scientific library (González-González, 2011) containing several incunabula, and a rich collection of historical scientific instruments (González-González, 1995).

## 2. Data

We have revised the boxes of documentation of the historical archive of the ROA related to astronomical observations following the classification established in the archive (González-González, 2010). The boxes consulted were: *i*) boxes 131-143 (Section: astronomy; subseries: Observations), *ii*) boxes 274-306 (Section: astronomy; subseries: Bounded observations), *iii*) boxes 369-386 (Section: astronomy; subseries: Observations), *iv*) boxes 1151-1201 (Section: astronomy; subseries: Observations), and *v*) boxes 1341-1369 (Section: astronomy; subseries: Observations). These boxes would form a column approximately 15 metres high if we stacked all the documentation consulted.

From this documentation, 7284 individual determinations of the solar radius have been recovered. These observations are not continuous but are distributed more or less throughout the period considered. After the treatment (see Section 3), the resulting annual values (Figure 1) form different clusters depending mainly on the instrument used (see the different colours in Figure 1). We recovered the measurements of the transit time of the Sun crossing the meridian of the Observatory, except for the last stage (when a Danjon astrolabe was used), in which measurements of solar radius were recovered directly.

In the following sub-sections, we will review what is known about the instruments, observers, and related information about these measurements of solar radius.

## 2.1. Mural Quadrant Manufactured by Bird

The most representative instrument of the astronomy of the 18th century was the mural quadrant. The main instrument of the ROA in its early observations was a six-foot mural quadrant manufactured by John Bird (London). This instrument was a twin of the quadrant of the Göttingen Observatory ($D$=0.05 m, $f$=1.86 m), which was also used by Tobias Mayer for determining the solar radius (Wittmann, 1980, 1988). The Observatory also had a transit telescope and a fine collection of scientific instruments of high quality (Lafuente and Sellés, 1988).

Using the mural quadrant, the first systematic program of astronomical observations was started in Cádiz by Vicente Tofiño and Josef Varela. They were also responsible for the installation arrangements and the adjustments to the mural. Two volumes of these observations were published (Tofiño and Varela, 1776, 1777) corresponding to the years 1773-1776. These observations are coloured purple in Figure 1.

The Spanish state would assign the Observatory a hydrographic orientation, and these first systematic observations ceased in 1777. Systematic observations were made again in the late 1780s. However, the observers found that the accuracy of the instrument had deteriorated, probably because of stability problems of the building and the lack of local instrumentalists for its maintenance (see Chapter VII, Section 2 of Lafuente and Sellés, 1988). Finally, the Observatory was transferred to San Fernando, a small town ten km away, where a new building was constructed (Figure 2). These observations are coloured blue in Figure 1.

The purchase of master instruments for the new building was made in the early 1830s. Therefore, the astronomical observations made during the first years of 19th century in San Fernando were made with small, portable, or provisionally installed instruments. These observations are coloured green in Figure 1. We have found in the ROA catalogue of instruments (González-González, 1995) several small astronomical quadrants that could have been used during the period 1801–1808. By comparison with the Bird mural quadrant, we estimate that the typical diameter of their objectives is approximately one cm.

**2.2. Meridian Telescope Manufactured by Jones**

New master instruments were finally installed in the ROA in 1833 and a new long series of observations started. The ROA historical archive has preserved a collection of manuscript logbooks containing all of the observations made with the meridian telescope ($D$=0.125 m and $f$=3.05 m) manufactured by Thomas Jones, following the meridian telescopes installed in the Greenwich and Kensington observatories. It was installed in the ROA in early 1833, and was in use until 1862. Figure 3 is an early photograph of this instrument. These observations are coloured yellow in Figure 1.

It is worth mentioning that these solar observations have been used in a different context. Vaquero and Gallego (2014) presented some marginal notes about sunspots that had been made during the period 1833–1840 in the manuscripts of the meridian solar observations preserved in the historical archive of the ROA. It must also be noted that the observations corresponding to the years 1833, 1834, and 1835 were published soon afterwards in three volumes (Sánchez Cerquero, 1835, 1836a, 1836b).

## 2.3. Meridian Circle Manufactured by Troughton and Simms

In 1856, the ROA director requested the purchase of new astronomical instruments. The meridian circle, a new type of master instrument, had emerged in the 19th century as the combination of transit telescopes and mural instruments as it allowed making determinations together (right ascension and declination) that other instruments made separately before. The meridian circle that was installed in the ROA in 1863 ($D$=0.203 m and $f$=3.530 m) was manufactured by Troughton and Simms (London) (Figure 4). Only a small fraction of these observations were published. The astronomical observations made in the years 1892 and 1893 were printed in the Annals of the Observatory (ROA, 1896, 1899).

We have recovered another long period of meridian solar observations made from 1892 to 1930 with the meridian circle manufactured by Troughton and Simms. However, some of the documents containing the original observation are lost, especially for the period 1863–1891, in which there are several important gaps. These observations are coloured red in Figure 1.

## 2.4. Danjon Astrolabe

The ROA made stellar observations with a Danjon astrolabe from 1968 to 1983. The purpose of these observations was to make measurements of the parameters of the Earth's rotation. When other techniques arose that were more useful to obtain these parameters, other uses were developed for Danjon astrolabes. In particular, the use of these instruments to monitor the solar radius was established (see, for example, Nollet, 2005). The ROA's Danjon astrolabe was also modified (Figure 5, Sánchez, 1991), establishing a program for the determination of the solar radius using this apparatus that lasted from 1991 to 2006 when the last measurements of the solar radius were made in the ROA.

The main results of the first campaigns were published (Sánchez *et al*., 1993, 1995). However, the rest of the measurements were not published, and we have recovered them from the ROA archive. These observations are coloured black in Figure 1.

## 3. Results

We used the complete set of periodic terms of the planetary VSOP87 (version D) theory (Bretagnon and Francou, 1988) to generate the solar coordinates (heliocentric ecliptic spherical coordinates for the equinox of the day). From the solar coordinates obtained with the VSOP87D theory, we computed the apparent declination and right ascension of the Sun using the standard procedures set out by Meeus (1998) and Seidelmann (2006). In our code, we used the following astronomical coordinates for Cádiz and San Fernando, respectively: 36º 32' N, 6º 17' W, 0 m, and 36º 28' N, 6º 12' W, 30 m. To test the code, we compared our data with those of the programs provided by the Institut de Mécanique Céleste et de Calcul des Éphémérides (IMCCE) (www.imcce.fr/en/ephemerides/) and the Jet Propulsion Laboratory (JPL-Horizons) (ssd.jpl.nasa.gov/?ephemerides), and the agreement was very good. The worst observable that we had was the time derivative of the right ascension, for which the discrepancies were below 1 % when confronting the data with the results of the Horizons program. We used the values of $\Delta T$ provided by the Horizons program (Giorgini *et al*., 1996). They vary from 16 seconds in 1773 to 65 seconds in 2006, presenting a minimum at 1893 ($\Delta T$=-7 s). These values of the $\Delta T$ parameter agree well those provided by Espenak and Meeus (eclipse.gsfc.nasa.gov/SEpubs/5MKSE.html). Moreover, these small discrepancies have little effect on our final results.

In order to convert the values of solar transit time recovered from the ROA archive to values of solar radius, we used the equation proposed by Wittmann and Neckel (1996). These values do not take into account corrections regarding to atmospheric refraction, diffraction at the objective of the instrument, personal equation of the observer, and the turbulence of the atmosphere (except for the modern observations with the Danjon Astrolabe published by ROA, which we assume they were corrected). To correct the rest of the values, we have followed the methodology of Toulmonde (1997). Thus, the final value for the semidiameter can be calculated by the expression: $R_c = R1 + Ref - \beta$, where $R_c$ is the semidiameter corrected, $R1$ is the semidiameter obtained from transit time, *Ref* is the correction for refraction, and *β* is the correction for diffraction (Table 1). In particular, following the methodology of Toulmonde (1997), we have used a mean value equal to 0.3" for the refraction correction. The diffraction correction is well approximated by the equation 1.22$\lambda$/*D*, where $\lambda$ represents the wavelength of the observations and *D* the diameter of the objective. This expression corresponds, in arcsec, approximately to 15"/*D*, *D* being the diameter of the objective in cm. The main difference among Toulmonde (1997) and our work is that he sometimes use a formula to estimate the diameter of the objective of the instruments, while we know in a more exact way the diameter of each instrument used in ROA. Moreover, he applies a constant diffraction correction of 0.7'' from 1870 and we correct all data according the diameter of the objective of the instrument used. Note that both refraction and diffraction corrections are applied to all historical measurements. We have not applied these corrections to the values obtained using the Danjon astrolabe because we have recovered the values corrected by ROA staff (Sánchez *et al.*, 1993; 1995).

We show in Figure 6 the annual averaged values of the solar radius (with 1$\sigma$ error bars) from the period 1773–2006. Furthermore, Figure 6 also shows two dashed lines that represent the interval 959.63"±3.00". The value 959.63" corresponds to the generally accepted value of solar semidiameter since Auwers (1891). Several works have evaluated the influence of different parameters such as personal equation (up to 1.5"), seeing (≈0.5"), and irradiation (≈1.5") on the measurements of solar radius, obtaining different results (Cullen, 1926; Gething, 1955; Wittmann, 1977; Wittmann, Bonet Navarro, and Wöhl, 1981; Smith and Messina, 1982). Thus, we have included these effects in the uncertainty (3.00").

## 4. Discussion

Following with the data plotted in Figure 6, we can study the possible long-term trend. To do that we have computed the least-squares regression line (marked by a red line in Figure 6). It can be seen that, taking into account the 1$\sigma$ error bars, all of the annual values of the solar radius are statistically compatible with the present value given by 959.63"±3.00". Finally, we report that the mean value of the solar radius calculated from the observations registered in the ROA during the last quarter of millennium (1773–2006) is equal to $R_{ROA}$ = 958.87"±1.77". This value is corrected by refraction and diffraction, and the uncertainty is represented by one standard deviation. If we compare our result with solar-radius data obtained by other authors (Auwers, 1891; Gething, 1955; Wittmann, 1977; Laclare *et al*., 1996; Toulmonde, 1997; Noël, 2005), it can be seen that the solar semidiameter from ROA is among the lowest values reported in the literature.

It can also be seen in Figure 6 that the higher values of the solar radius, as well as the greatest uncertainties, correspond to the measurements made in the periods 1788–1790 and 1801–1808. This fact can be related to the problems of maintenance and stability of the building during 1788–1790 and the use of small, portable, and less accurate instruments in the earlier years of 19th century.

The analysis of the least-squares fit ($R^2$ = 0.0001363; $F$ = 0.008995; $p$-value = 0.9247) shows no significant trend, whether increasing or decreasing, in the solar radius. The main conclusion of this statistical analysis is the presence of no observable trend in the ROA measurements of the solar radius (see Figure 6).

We have finally performed a comparison between the values of the solar radius obtained from the measurements in the first stage of the ROA (1773–1776) and the values retrieved from the observations made by Tobias Mayer at the Göttingen Observatory during the period 1756–1760 (Wittmann, 1980). These observations were made using two similar mural quadrants constructed by John Bird. We have not included the measurements of the second stage at ROA (which were made with the same instrument) due to the problems reported by the observers. Table 2 presents the annual average values for solar radius at the ROA and Göttingen Observatory. In Table 2, one can see that the corrected values of the solar semidiameter obtained for ROA (957.24"±0.69") and Göttingen (957.40"±0.41") with Bird mural quadrants are very close. Additionally, these data are within the interval 959.63"±3.00". Note that the relatively low values obtained with these twin mural quadrants may be related to this kind of instrument. The RHtestV4 test (Wang and Feng, 2013) was used to test the homogeneity of the annual solar radius data from ROA (first stage) and Göttingen Observatory made with a similar Bird mural quadrant. This test is based on the penalized maximal $t$ and the penalized maximal $F$-tests embedded in a recursive testing algorithm. The results showed no change point exists in this series and, therefore, these data can be considered homogeneous.

In order to analyse the connection between solar radius and solar activity, we have compared the values of the solar semi-diameter obtained from measurements in the ROA with the index recently published of solar activity, sunspot number (Clette *et al*., 2015). The values of sunspot number index have been extracted from www.sidc.be/silso/datafiles. In Figure 7 (upper panel), we represent the annual value of the solar radius from ROA (1773–2006) versus sunspot-number index. The correlation coefficient between the two datasets is equal to $r$=0.091. According to ROA data, this low correlation implies that the variations of the solar radius and sunspot number do not have any relationship. Moreover, Figure 7 (bottom panel) shows the temporal evolution of solar semi-diameter (black dots) from the Jones meridian telescope for the fourth stage at ROA (1833–1868) and sunspot number (red line) around that time. We note that the error bars represent two standard deviations and the value for 1866 does not have an error bar since we have only one record. Although this stage presents the greatest correlation between solar radius and solar activity for all stages at ROA, the correlation is also low. In this case, the correlation coefficient is equal to $r$=0.109. In any case, it can be seen that the correlation is very low and, therefore, no relationship between solar radius and sunspot number exists from the measurements made in ROA during the period 1773–2006. This result agrees with the previous studies by Bush, Emilio, and Kuhn (2010). They used the Michelson Doppler Imager (MDI) onboard the Solar and Heliospheric Observatory (SOHO) spacecraft to obtain the first homogeneous and highly precise measurements of the solar radius over a complete solar cycle, showing that these variations during a solar cycle are very small.

## 5. Conclusions

The ROA was founded in 1753. This Spanish institution has spent more than 260 years making astronomical observations almost continuously. The rich documentation in its archive has allowed us to rescue a remarkable set of solar-radius measurements made throughout the history of the ROA. Furthermore, we recovered much useful metadata associated with these observations. We highlighted the different instruments that were used in this type of observation: a Bird mural quadrant, Jones meridian telescope, Troughton and Simms meridian circle, and a modified Danjon solar astrolabe.

We have undertaken a time-consuming task to locate and recover a total of 7284 reliable individual determinations of the solar radius covering the period 1773–2006. Thus, a new collection of thousands of records of solar semi-diameter has been recovered to study the solar diameter and its long-term variations. Six different stages in the history of observations can be established (Figure 1). In the second and third stages, the values of the solar radius are excessively high. We think that these anomalous data were due to stability problems of the old building of the observatory in Cadiz (stage 2) and the use of small, or portable or provisionally installed, instruments (stage 3).

Other values also presented significant differences depending on the instrument being used. Thus, the values of the solar radius obtained with the Troughton and Simms Meridian circle were the highest of the series.

From a long-term perspective, these data do not present any significant trend from the statistical point of view. If this long-term trend existed, it would be within the measurement uncertainty. Therefore, these solar observations are evidence for no change in the solar diameter in the last quarter of a millennium. This result agrees with the previous studies of the long-term variation of the solar radius (Toulmonde, 1997; Rozelot, 2001) that show no trend since the middle of the 18th century. There is some uncertainty in the values of the solar radius during the Maunder Minimum (1645–1715), an epoch of very low solar activity (Eddy, 1976; Usoskin *et al*., 2016). Further studies should be conducted recovering ancient records of the solar radius to resolve this issue. Moreover, after comparing the solar radius calculated from ROA and sunspot-number index, we have not found any relationship between solar radius and sunspot number. Finally, we have calculated the mean value for the solar radius from the measurements made in the ROA during the period 1773–2006 including the corrections by diffraction and refraction. Thus, we have obtained: $R_{ROA} = 958.87"\pm1.77"$.

**Acknowledgements**

Support from the Junta de Extremadura (Research Group Grants GR15137 and GRU10158) and from the Ministerio de Economía y Competitividad of the Spanish Government (FIS2013-42840-P, AYA2011-25945, and AYA2014-57556-P) is gratefully acknowledged. We thank the referee for helpful comments and suggestions.

**Disclosure of Potential Conflicts of Interest**

The authors declare that they have no conflicts of interest.

Table 1. Solar radius observations made at the ROA during 1773–2006. We show i) the year of the observations, ii) the value of the solar radius from transit times without corrections, iii) the corrected value by refraction and diffraction of the solar radius, iv) the standard deviation of the corrected radius and v) the instrument used (BQ – Bird mural quadrant; SP – small and portable instruments; JT – Jones meridian telescope; TS - Troughton and Simms meridian circle; DA – Danjon astrolabe). Note that the historical solar radius measurements are corrected for refraction and diffraction. We have not applied these corrections to the values obtained using the Danjon astrolabe because we have recovered the values corrected by ROA staff (Sánchez *et al*., 1993; 1995).

| Year | Uncorrected Radius | Corrected Radius | Standard Deviation | Instrument |
|---|---|---|---|---|
| 1773 | 960.42 | 957.72 | 3.97 | BQ |
| 1774 | 959.87 | 957.17 | 2.30 | BQ |
| 1775 | 959.84 | 957.14 | 1.87 | BQ |
| 1776 | 959.62 | 956.92 | 2.47 | BQ |
| 1788 | 967.47 | 964.77 | 6.85 | BQ |
| 1789 | 964.33 | 961.63 | 5.35 | BQ |
| 1790 | 963.83 | 961.13 | 4.23 | BQ |
| 1801 | 977.36 | 962.66 | 3.56 | SP |
| 1802 | 972.08 | 957.38 | 0.47 | SP |
| 1804 | 976.50 | 961.80 | 4.81 | SP |
| 1805 | 976.26 | 961.56 | 4.62 | SP |
| 1806 | 977.50 | 962.80 | 4.57 | SP |
| 1807 | 977.21 | 962.51 | 2.79 | SP |
| 1808 | 977.33 | 962.63 | 2.90 | SP |
| 1833 | 958.81 | 957.91 | 1.56 | JT |
| 1834 | 958.75 | 957.85 | 0.89 | JT |
| 1835 | 959.10 | 958.20 | 1.12 | JT |
| 1836 | 959.68 | 958.78 | 1.06 | JT |
| 1837 | 958.39 | 957.49 | 1.37 | JT |
| 1838 | 958.27 | 957.37 | 1.88 | JT |
| 1839 | 957.78 | 956.88 | 1.66 | JT |
| 1840 | 957.89 | 956.99 | 1.30 | JT |
| 1841 | 957.75 | 956.85 | 1.22 | JT |
| 1842 | 957.82 | 956.92 | 1.42 | JT |
| 1843 | 958.27 | 957.37 | 1.14 | JT |
| 1844 | 958.13 | 957.23 | 0.99 | JT |
| 1845 | 957.44 | 956.54 | 0.93 | JT |
| 1846 | 957.62 | 956.72 | 0.91 | JT |
| 1847 | 957.98 | 957.08 | 1.16 | JT |
| 1848 | 958.16 | 957.26 | 1.44 | JT |
| 1849 | 958.54 | 957.64 | 1.23 | JT |
| 1850 | 958.92 | 958.02 | 1.43 | JT |
| 1851 | 959.06 | 958.16 | 1.25 | JT |
| 1852 | 959.28 | 958.38 | 1.77 | JT |
| 1853 | 959.26 | 958.36 | 1.23 | JT |
| 1854 | 958.97 | 958.07 | 0.97 | JT |
| 1855 | 958.28 | 957.38 | 0.67 | JT |
| 1856 | 958.88 | 957.98 | 0.84 | JT |
| 1857 | 959.31 | 958.41 | 0.85 | JT |
| 1858 | 959.63 | 958.73 | 0.96 | JT |
| 1859 | 959.43 | 958.53 | 0.86 | JT |
| 1866 | 957.61 | 956.71 | -    | JT |
| 1867 | 958.36 | 957.46 | 1.12 | JT |
| 1868 | 958.78 | 957.88 | 1.40 | JT |
| 1892 | 960.66 | 960.23 | 1.10 | TS |
| 1893 | 960.90 | 960.46 | 1.08 | TS |
| 1906 | 960.36 | 959.92 | 1.42 | TS |
| 1907 | 960.60 | 960.16 | 1.34 | TS |
| 1908 | 960.39 | 959.96 | 1.42 | TS |
| 1909 | 960.23 | 959.79 | 1.18 | TS |
| 1910 | 960.19 | 959.75 | 1.35 | TS |
| 1911 | 960.55 | 960.11 | 2.24 | TS |
| 1928 | 960.53 | 960.09 | 1.44 | TS |
| 1929 | 960.65 | 960.21 | 0.47 | TS |
| 1930 | 960.69 | 960.25 | 2.40 | TS |
| 1991 | 958.59 | 958.59 | 0.71 | DA |
| 1992 | 958.54 | 958.54 | 0.73 | DA |
| 1993 | 959.31 | 959.31 | 1.09 | DA |
| 1994 | 959.18 | 959.18 | 1.16 | DA |
| 1998 | 959.33 | 959.33 | 0.45 | DA |
| 1999 | 958.98 | 958.98 | 0.39 | DA |
| 2000 | 958.98 | 958.98 | 0.37 | DA |
| 2001 | 959.03 | 959.03 | 0.36 | DA |
| 2002 | 958.75 | 958.75 | 0.49 | DA |
| 2003 | 958.67 | 958.67 | 0.49 | DA |
| 2004 | 958.46 | 958.46 | 0.51 | DA |
| 2005 | 958.83 | 958.83 | 0.48 | DA |
| 2006 | 958.85 | 958.85 | 0.75 | DA |

Table 2. Annual corrected values for solar semidiameter measured at the Göttingen Observatory (1756–1760) and ROA (1773–1776). These observations were made with similar instruments: two Bird mural quadrants. The uncertainties are given by two standard deviations.

| Observatory | Date | Solar Radius |
|---|---|---|
| Göttingen | 1756 | 957.53 |
| | 1757 | 957.13 |
| | 1758 | 957.58 |
| | 1760 | 957.37 |
| | 1756 – 1760 | 957.40 ± 0.41 |
| ROA | 1773 | 957.72 |
| | 1774 | 957.17 |
| | 1775 | 957.14 |
| | 1776 | 956.92 |
| | 1773 – 1776 | 957.24 ± 0.69 |

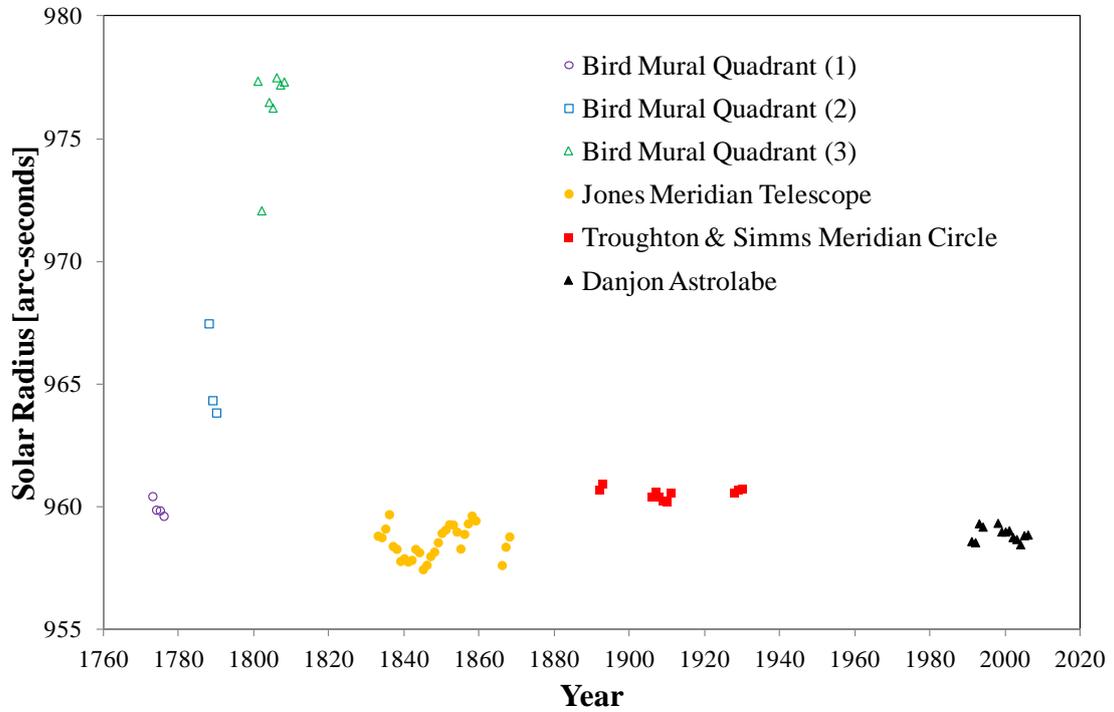

Figure 1. Annual values of the solar radius in arc-seconds from measurements made at the ROA during approximately the last quarter of a millennium.

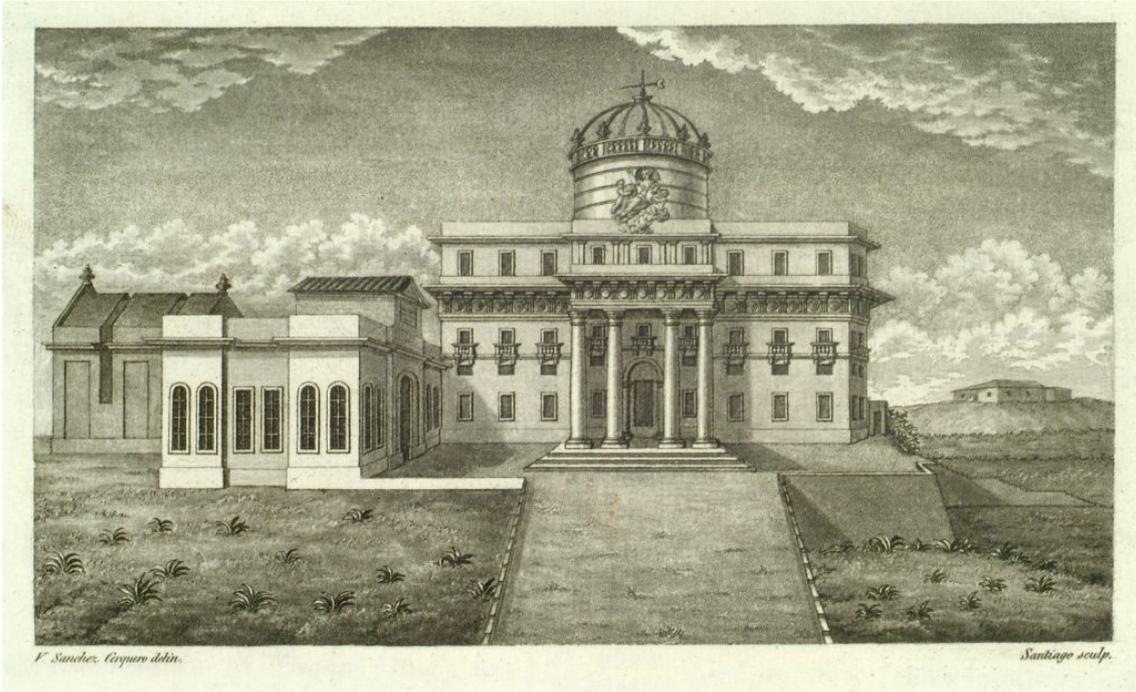

Figure 2. An engraving of the ROA building around 1830 (courtesy of the ROA Archive).

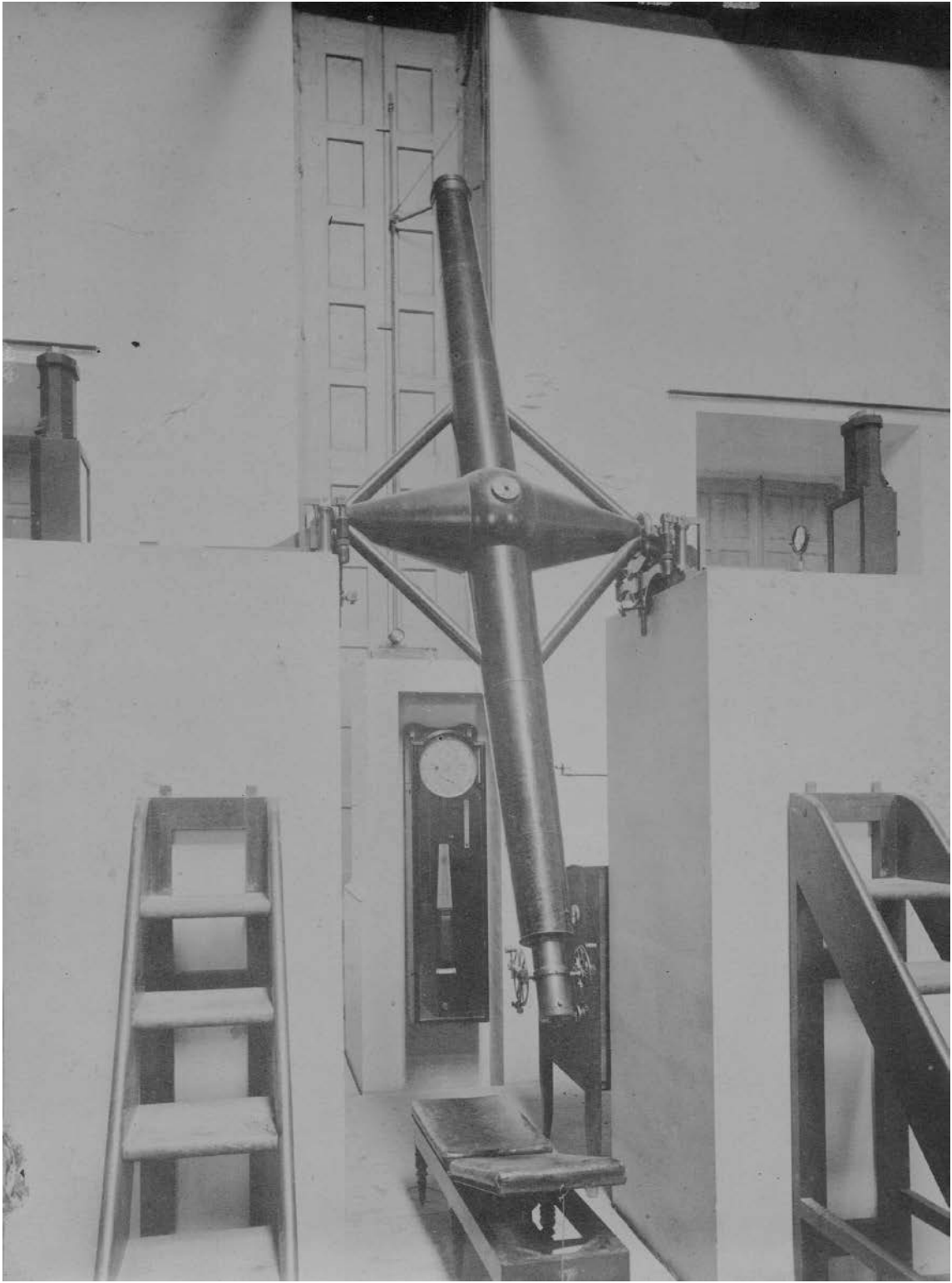

Figure 3. A historical photograph of the meridian telescope manufactured by Thomas Jones and installed in the ROA (courtesy of the ROA Archive).

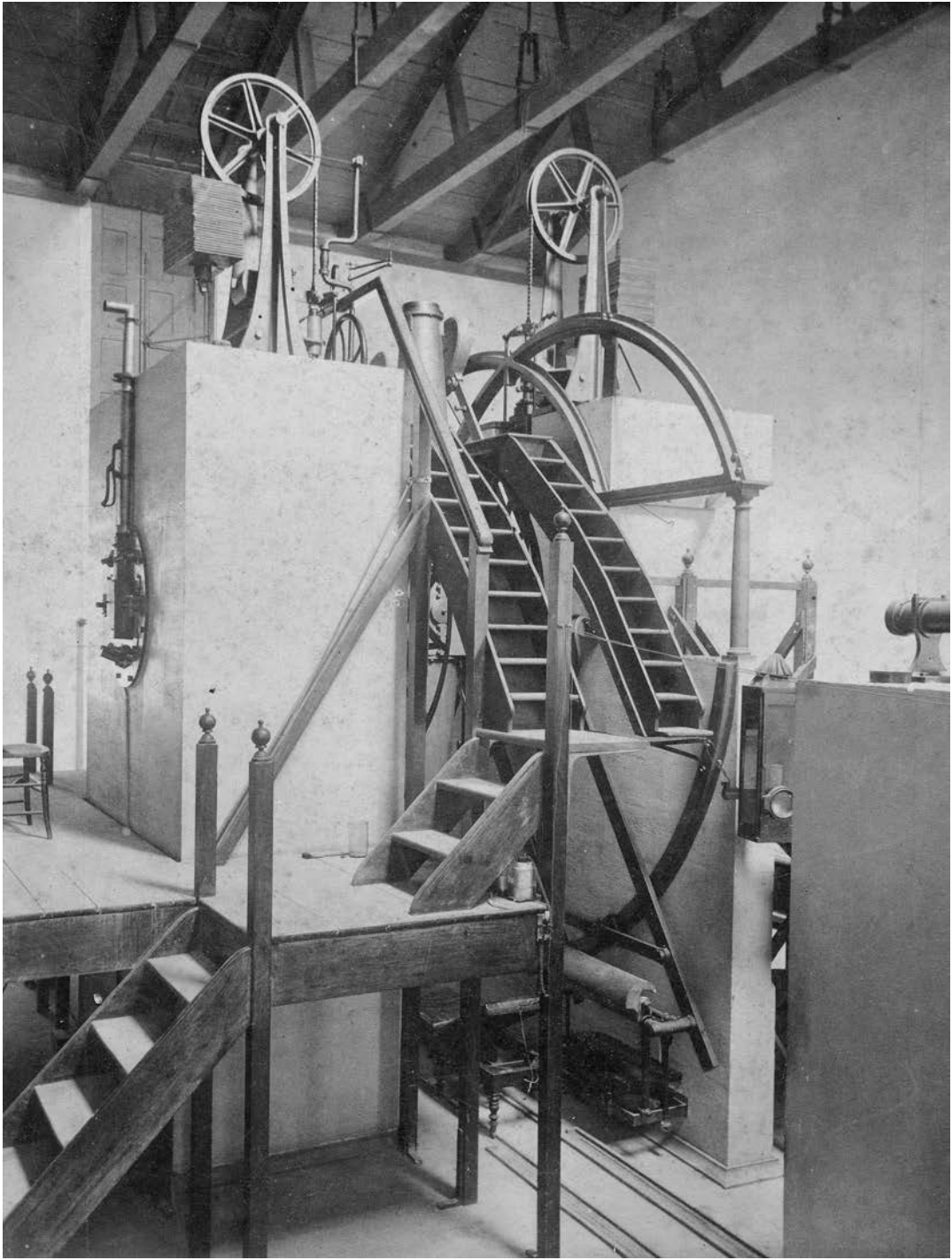

Figure 4. A historical photograph of the meridian telescope manufactured by Troughton and Simms installed at the ROA (courtesy of the ROA Archive).

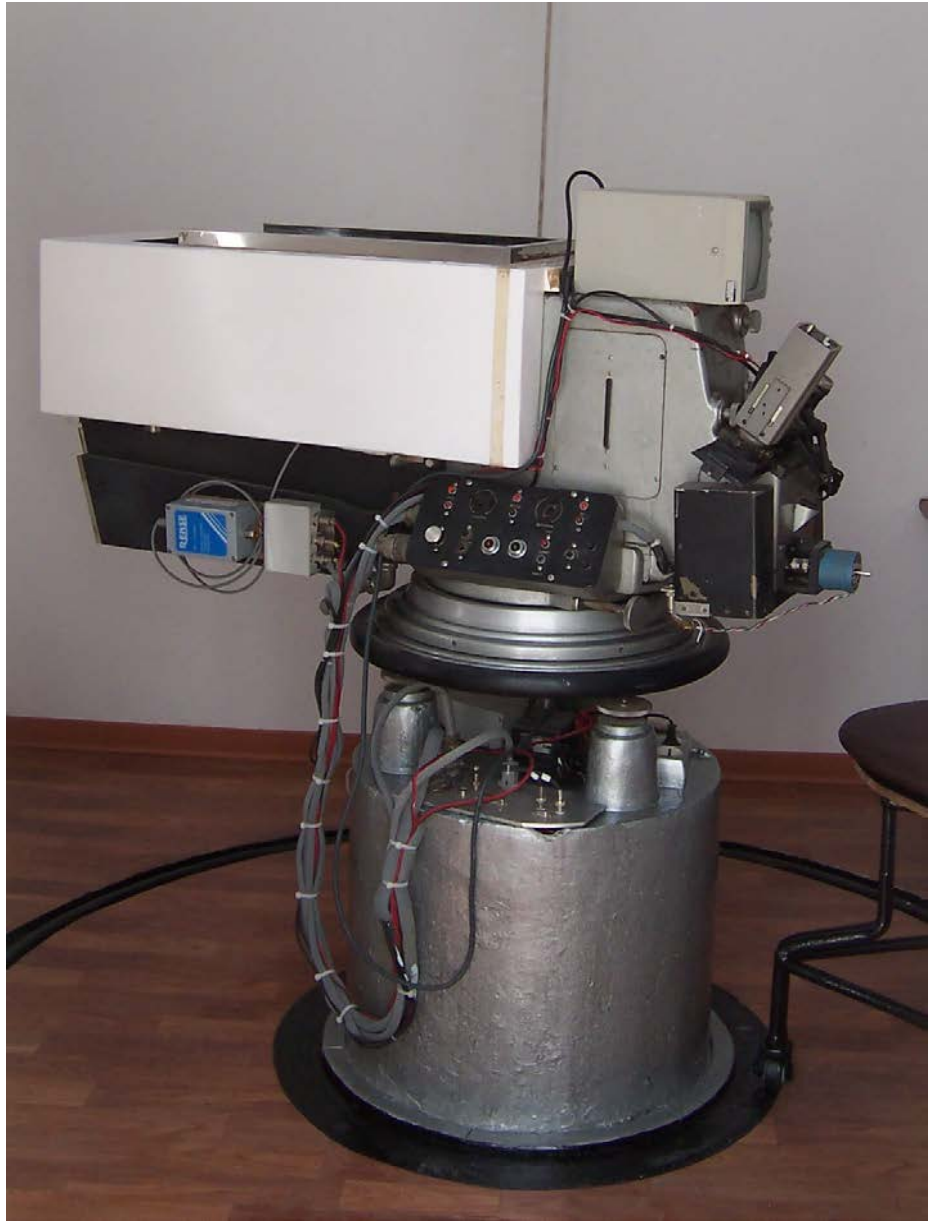

Figure 5. Modified Danjon astrolobe (courtesy of the ROA archive).

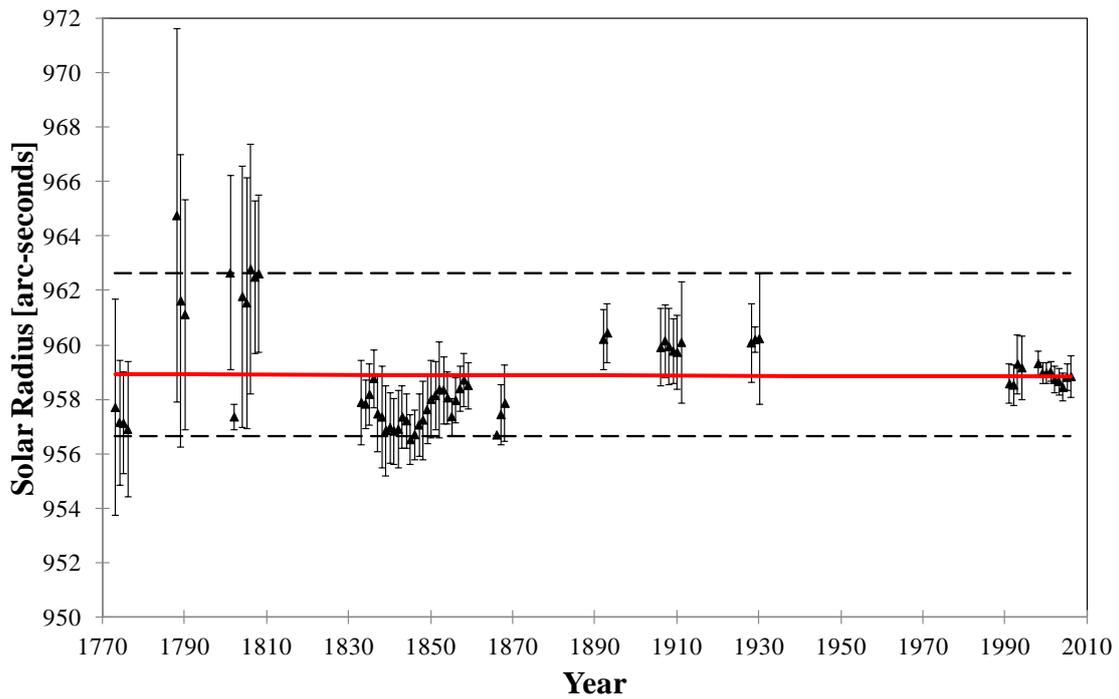

Figure 6. Annual averaged values of the measurements of the solar radius in arc-seconds with error bars (1$\sigma$) made at the ROA in Spain during approximately the last quarter of a millennium. Heavy line is the least-squares fit. The two horizontal dashed lines represent the interval 959.63"±3.00".

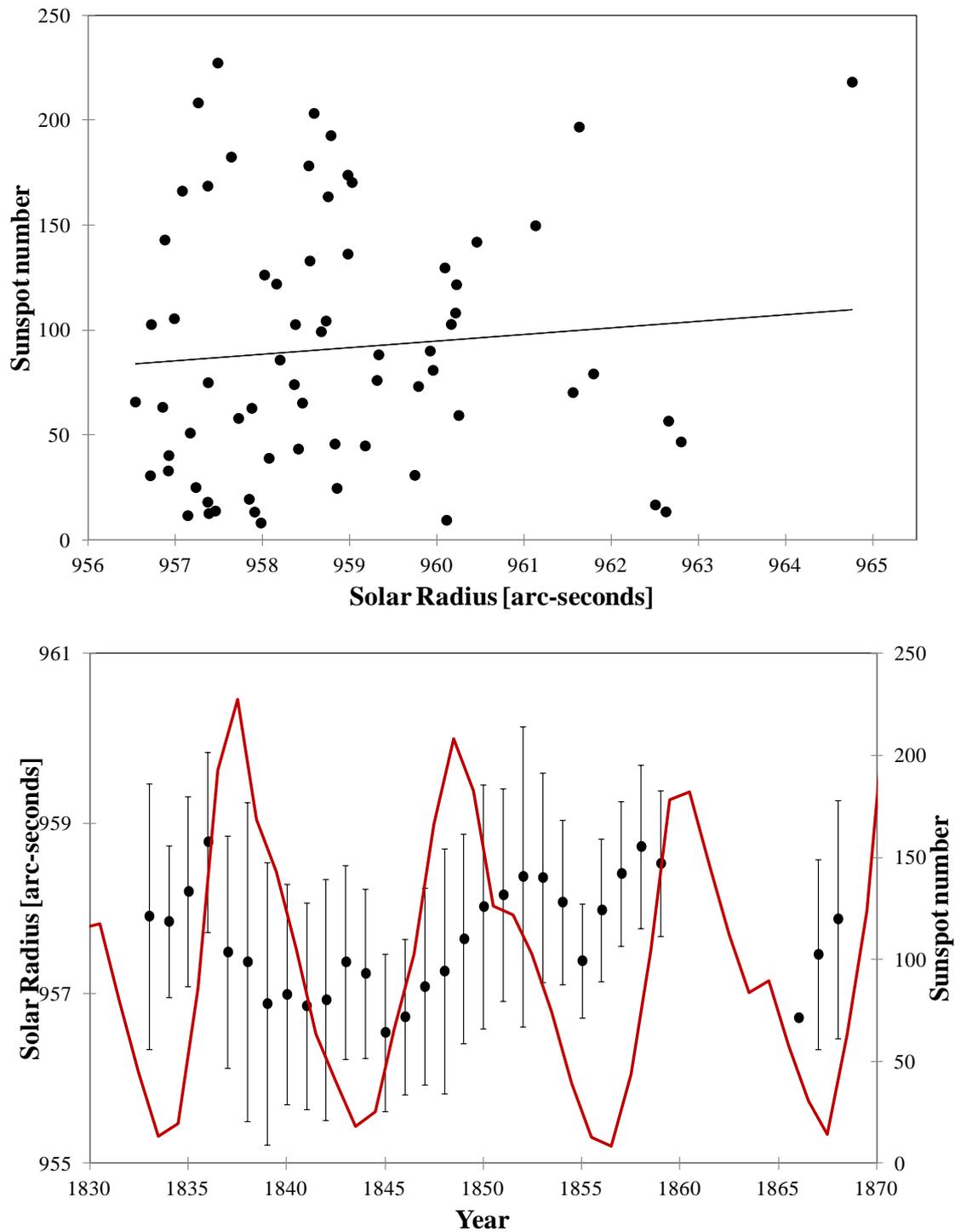

Figure **7**. Linear relationship between solar radius from ROA and sunspot number (top) and temporal evolution of solar radius in ROA during 1833–1868 (black dots) and sunspot number (heavy line). Error bars represent one standard deviation. [Sunspot number source: WDC-SILSO, Royal Observatory of Belgium, Brussels]